\documentclass[final,5p,times,twocolumn]{elsarticle}
\usepackage{graphicx}
\usepackage{amsmath}
\usepackage{geometry}
\usepackage{amssymb}
\biboptions{compress}

\journal{A Journal}

\begin{document}
	
\begin{frontmatter}

\title{Accurate expansion of cylindrical paraxial waves \\ for its straightforward implementation in electromagnetic scattering}

\author[label1,label2]{Mahin Naserpour}
\author[label1]{Carlos J. Zapata-Rodr\'{i}guez\corref{cor*}}
\cortext[cor*]{Corresponding author:}
\ead{carlos.zapata@uv.es}
\address[label1]{Department of Optics and Optometry and Vision Science, University of Valencia, Dr. Moliner 50, Burjassot 46100, Spain}
\address[label2]{Physics Department, College of Sciences, Shiraz University, Shiraz 71946-84795, Iran}

\begin{abstract}
The evaluation of vector wave fields can be accurately performed by means of diffraction integrals, differential equations and also series expansions.
In this paper, a Bessel series expansion which basis relies on the exact solution of the Helmholtz equation in cylindrical coordinates is theoretically developed for the straightforward yet accurate description of low-numerical-aperture focal waves.
The validity of this approach is confirmed by explicit application to Gaussian beams and apertured focused fields in the paraxial regime.
Finally we discuss how our procedure can be favorably implemented in scattering problems.
\end{abstract}

\begin{keyword}
	Focal waves \sep diffraction \sep wave expansion.
\end{keyword}

\end{frontmatter}

\section{Introduction}

The description of unbounded wave fields in the form of series expansion represents a widely used tool in free space propagation \cite{Stratton41,Jackson75,Siegman86}. 
For instance, the electric field emerging from a laser device is commonly expressed as a combination of either Hermite-Gauss or Laguerre-Gauss basis functions taken from the natural modes existing in mirror cavities \cite{Kogelnik66,Allen92,Saghafi98}. 
Such procedure, however, seems inappropriate for apertured focal waves where diffraction sidelobes will impose a strong limitation in its efficient characterization by an acceptable truncation of the sequence \cite{Erkkila81,Stamnes86}. 
In these cases, expansions in terms of Lommel functions demonstrate a suitable implementation \cite{Born99,Sheppard14}.

In planar arrangements, two-dimensional (2D) waves satisfying the Helmholtz equation can be described in terms of Bessel wave functions when they are formulated in cylindrical coordinates \cite{Balanis89}. 
This is a method followed for instance in the Mie-Lorenz theory applied to scatterers with circular cross section \cite{Shah70,Bohren98,Diaz16c}.
Only a small part of the cases analyzed in scattering problems consider non-uniform beams, among other reasons due to the complexity of the series expansions with cylindrical vector wave functions (also spherical vector wave functions) that can be found \cite{Kojima79,Zimmermann95,Gouesbet97,Zhang07,Pawliuk09,Wang17}.
Particularly interesting is the paper from Shogo Kozaki which analyzes the case of a Gaussian beam illuminating a cylindrical particle, where it is possible to remarkably simplify the analytical description of the beam, under certain approximations, in terms of Bessel cylindrical waves \cite{Kozaki82}.

A salient feature of these Bessel wave functions is their inherent localization around the chosen origin of coordinates. 
Therefore, focal waves are generally expected to be favorable candidates to be effectively represented by means of a series expansion using Bessel wave fields.

In this paper, we consider 2D paraxial wave fields which are localized in the vicinity of a given point, which serves for the construction of the Bessel wave-functions basis to be used in a series expansion. 
The coefficients of such sequence are estimated by analyzing the far field of the paraxial wave. 
This choice reduces the resultant calculation to a simple Fourier expansion. 
The validity of this method is finally verified when applied to a Gaussian laser beam and also to an apertured focal wave.

\section{Theoretical analysis}

Let us consider a 2D wave field $E(x,z) = \exp(i k z) U(x,z)$ propagating in free space and satisfying the paraxial wave equation, $\partial_x^2 U + 2 i k \partial_z U = 0$, where $z$ represents the spatial coordinate along the optical axis, $x$ is the transverse coordinate and $k = 2 \pi / \lambda$ is the wavenumber.
Note that a harmonic time variation $\exp(-i \omega t)$ is here assumed, where $\omega$ is the time-domain frequency.
If the focal point is set at the origin of coordinates, $(x,z) = (0,0)$, the wave field at any out-of-focus plane $z \neq 0$ can be determined by means of the Fresnel diffraction integral \cite{Sheppard13}
\begin{equation}
 E(x,z) = \frac{\exp(i k z)}{\sqrt{i \lambda z}} \int_{-\infty}^{\infty} E(x_0,0) \exp \left[ \frac{i k}{2 z} \left( x - x_0 \right)^2 \right] d x_0 .
 \label{eq01}
\end{equation}
The characteristic Fraunhofer pattern can be observed well beyond the focal plane in the limit $|z| \to \infty$.
In this case, the quadratic phase term $\exp \left( {i k} x_0^2 / {2 z} \right)$ in the integral equation (\ref{eq01}) is negligible.

When $z>0$, the far field of $E(x,z)$ can be expressed as
\begin{equation}
 E_F^+(x,z) = \frac{\exp(i k r)}{\sqrt{r}} \exp(-i \pi/4) \sqrt{\lambda} a_F (\theta) ,
 \label{eq02}
\end{equation}
where the angular spectrum
\begin{equation}
 a_F (\theta) = \frac{1}{\lambda} \int_{\infty}^{\infty} E(x_0,0) \exp \left( -i k \theta x_0 \right) d x_0 ,
 \label{eq03}
\end{equation}
is simply the Fourier transform of the wave field at the focal plane.
The azimuthal angle measured from the optical axis is given as $\theta = x/z$ in the paraxial approximation, whereas the radial coordinate is given by $r = |z + x^2 / 2 z|$.
In the paraxial regime, the wave function $a(\theta)$ has significant values only when $\theta \ll 1$.
Furthermore, the focal field can be evaluated once the angular spectrum $a_F(\theta)$ is known as 
\begin{equation}
 E(x,z) = \exp (i k z) \int_{-\infty}^{\infty} a_F(\theta) \exp \left( -i \frac{k}{2} z \theta^2 \right) \exp \left( i k \theta x \right) d \theta .
 \label{eq14}
\end{equation}
This diffraction integral is well known within the Debye theory and has been successfully applied to apertured focal waves, as we will consider below \cite{Zapata00}.

Now it is clear that a wave expansion can be given by means of a Fourier series of the angular spectrum, that is 
\begin{equation}
 a_F(\theta) = \sum_{n = -\infty}^{\infty} a_n \exp(i n \theta) ,
 \label{eq04}
\end{equation}
where the Fourier coefficient
\begin{equation}
 a_n = \frac{1}{2 \pi} \int_{-\pi}^{\pi} a_F(\theta) \exp(-i n \theta) d \theta .
 \label{eq05}
\end{equation}
Since the angular spectrum $a(\theta)$ takes values identically zero in the range $|\theta| > \pi/2$ in the semi-space $z > 0$, the definite integral given in Eq.~(\ref{eq05}) can be further simplified by extending the interval of integration to $-\infty < \theta < \infty$.
Under such approximation, and substituting Eq.~\eqref{eq03} into \eqref{eq05}, we finally infer that the Fourier coefficient of order $n$  
\begin{equation}
 a_n = \frac{1}{2 \pi} E \left( - \frac{n}{k},0 \right) ,
 \label{eq11}
\end{equation}
depends on the focal field as measured at the off-axis point $x=-n/k$.

To extend the series expansion of the wave field into the near field (note that here we disregard evanescent waves and near field refers to waves in the vicinities of focus), we will use the Bessel wave functions of the first kind and order $n$, $J_n (k r)$, and the Bessel functions of the second kind and order $n$, $Y_n (k r)$, which are solutions of the 2D Helmholtz wave equation, $\nabla^2 E + k^2 E = 0$, provided that the angular variation of the wave field is given by $\exp(i n \theta)$ \cite{Balanis89}.
In particular, the Hankel wave function of the first kind $H_n^{(1)} (k r) = J_n (k r) + i Y_n (k r)$ has an asymptotic limit far from the origin of coordinates given as
\begin{equation}
 H_n^{(1)} (k r) \to \sqrt{\frac{2}{\pi k r}} \exp(i k r) \exp(-i n \pi / 2) \exp(-i \pi/4) .
 \label{eq06}
\end{equation}
Therefore, an outgoing cylindrical wave field with a specific angular momentum can be given in terms of a Hankel wave function of the first kind and unique order $n$.
As a consequence, a wave field exhibiting an angular spectrum $a_F(\theta)$ in the semi-space $z>0$ can be expressed as a combination of Hankel wave functions of different orders $n$ as
\begin{equation}
 E_F^+(x,z) = \pi \sum_{n = -\infty}^{\infty} a_n \exp(i n \pi / 2) H_n^{(1)} (k r) \exp(i n \theta) ,
 \label{eq07}
\end{equation}
which in addition can be utilized in the near field.

Equation (\ref{eq07}) provides the expression of a wave field whose Fraunhofer pattern is given in terms of the wave function $a_F(\theta)$, leading to a vanishing far field in $z < 0$.
However, in such semi-space, the far field of $E(x,z)$ given in Eq.~(\ref{eq01}) can be expressed as
\begin{equation}
E_F^-(x,z) = \frac{\exp(-i k r)}{\sqrt{r}} \exp(i \pi/4) \sqrt{\lambda} a_F (\theta') ,
\label{eq08}
\end{equation}
where the angle $\theta' = x/z$ in the paraxial approximation.
With the new angular coordinate $\theta' = \theta - \pi$, we again may describe the Fraunhofer pattern observed at $z \to - \infty$ by means of the wave function $a_F (\theta')$ which takes significant values for $|\theta'| \ll 1$.
Furthermore, the Fourier series given in Eqs.~(\ref{eq04}) and (\ref{eq05}) are still applicable in this case.
Taking into account the asymptotic limit of the Hankel wave function of the second kind $H_n^{(2)} (k r) = J_n (k r) - i Y_n (k r)$, written as 
\begin{equation}
H_n^{(2)} (k r) \to \sqrt{\frac{2}{\pi k r}} \exp(-i k r) \exp(i n \pi / 2) \exp(i \pi/4) ,
\label{eq09}
\end{equation}
we may infer a series expansion of the wave field in terms of incoming cylindrical waves, resulting in
\begin{equation}
E_F^-(x,z) = \pi \sum_{n = -\infty}^{\infty} a_n \exp(-i n \pi / 2) H_n^{(2)} (k r) \exp(i n \theta') .
\label{eq10}
\end{equation}

The appropriate description of the paraxial wave field $E(x,z)$ includes a far field at $z \to \infty$ in the form of an outgoing cylindrical wave patterned by $a_F(\theta)$, and simultaneously at $z \to -\infty$ representing an ingoing cylindrical wave shaped by the same angular spectrum $a_F(\theta')$.
Note that such symmetry around the focal point has been analyzed by Collet and Wolf \cite{Collet80}.
As a consequence, the wave field must be computed as $E(x,z) = E_F^+(x,z) + E_F^-(x,z)$, which in cylindrical coordinates is expressed as
\begin{equation}
E(r,\theta) = 2 \pi \sum_{n = -\infty}^{\infty} i^n a_n J_n (k r) \exp(i n \theta) .
\label{eq12}
\end{equation}
This is the main result of our study, together with the fact that the Fourier coefficients as given in Eq.~(\ref{eq11}) can be achieved in terms of the focal wave field.
We conclude that a cylindrical, paraxial wave field can be accurately described by a series expansion sustained by Bessel wave functions which are solutions of the 2D Helmholtz equation, and with expansion coefficients which are determined by the own wave field at specific points of the focal plane.
We point out that Eq.~(\ref{eq12}) accurately provides the wave field even at angles far from the paraxial regime were $\tan \theta = x/z$; of course $r = \sqrt{x^2 + z^2}$ denotes the distance from focus to the observation point. 

In centrosymmetric field distributions where $E(-x,z) = E(x,z)$, what occurs simply if the field is symmetric with respect to the origin of coordinates at the focal plane, a reduced expression of the series expansion can be deduced.
In this case, $a_{-n} = a_n$ as inferred from Eq.~(\ref{eq11}).
By using the following property of Bessel functions of negative order, $J_{-n} (\alpha) = (-1)^n J_n (\alpha)$, we finally may reduce Eq.~(\ref{eq12}) to
\begin{equation}
E(r,\theta) = E_0 J_0 (k r) + 4 \pi \sum_{n = 1}^{\infty} i^n a_n J_n (k r) \cos(n \theta) ,
\label{eq13}
\end{equation}
where $E_0 = 2 \pi a_0$ is the in-focus wave field.

\section{Implementation in scattering problems}

\begin{figure}[tb]
	\centering
	\includegraphics[width=.8\linewidth]{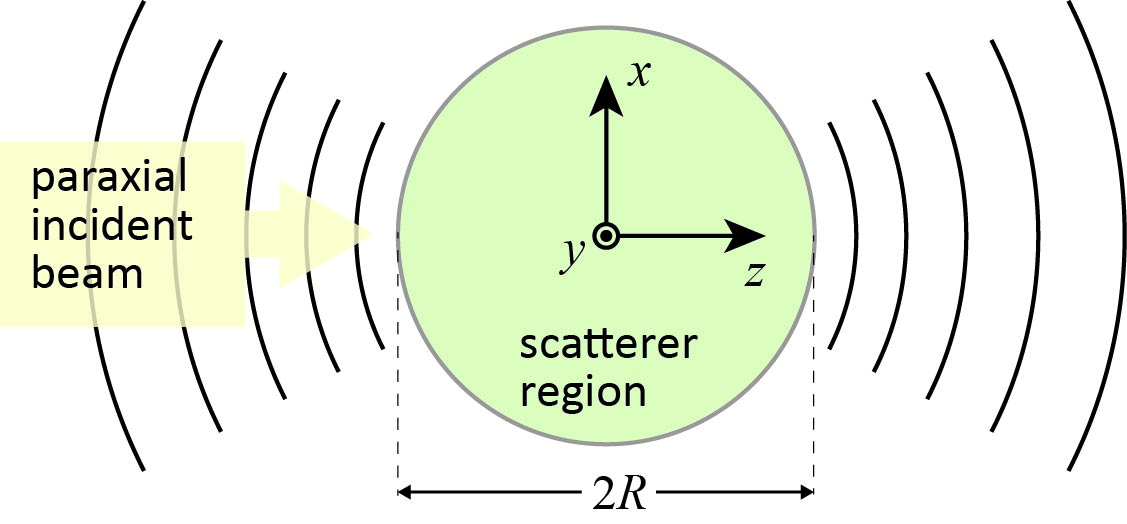}
	\caption{Illustration of the paraxial beam impinging on the region of interest where the scattering body is located.}
	\label{fig08}
\end{figure}

Let us consider a scatterer embodied in a cylindrical region whose axis is set along the $y$ axis and with a radius $R$, as depicted in Fig.~\ref{fig08}.
In order to analytically estimate the scattered wave field of the body, we may follow the Lorenz-Mie scattering method which is described in detail for instance in Refs.~\cite{Shah70,Bohren98,Bussey75}.
Here we consider TM$^y$ polarized waves in such a way that the electric field is oriented along the $y$ axis.
Note that the case of TE$^y$ polarized waves where the magnetic field lies on the $y$ axis is straightforwardly treated by using the duality theorem \cite{Balanis89}.
Using Eq.~(\ref{eq12}), the electromagnetic field of the incident wave field may be written as 
\begin{equation}
\mathbf{E}_{in} = \hat{z} E_0 \sum_{n = - \infty}^{+\infty} 2 \pi a_n i^n J_n \left(k r \right) \exp \left( i n \theta \right) ,
\end{equation}
and $\mathbf{H}_{in} = - i {\nabla \times \mathbf{E}_{in}}/{\omega \mu_0}$, which reads as
\begin{equation}
\mathbf{H}_{in} = H_0 \sum_{n=-\infty}^{+\infty} 2 \pi a_n i^n \left[ \hat{r} n \frac{J_n \left( kr \right)}{k r} 
+ \hat{\theta} i J'_n \left( kr \right) \right] \exp \left( i n \theta \right) ,
\end{equation}
where $H_0 = {E_0 k} / { \omega \mu_0}$ is a constant amplitude.
Here the prime appearing in $J'_n \left( \alpha \right)$ denotes derivative with respect to the variable $\alpha$.
The scattered electromagnetic field in the environment medium, $r > R$, may be set as \cite{Bohren98}
\begin{equation}
\mathbf{E}_{sca} = \hat{z} E_0 \sum_{n = - \infty}^{+\infty} b_n i^n H_n^{(1)} \left(k r \right) \exp \left( i n \theta \right) ,
\end{equation}
and
\begin{equation}
\mathbf{H}_{sca} = H_0 \sum_{n=-\infty}^{+\infty} b_n i^n \left[ \hat{r} n \frac{H_n^{(1)} \left( k r \right)}{k r}   
+ \hat{\theta} i {H_n^{(1)}}' \left( kr \right) \right] \exp \left( i n \theta \right),
\end{equation}
where the coefficients $b_n$ must be determined.
The total electric field in the environment medium is simply $\mathbf{E}_{tot} = \mathbf{E}_{in} + \mathbf{E}_{sca}$.

The Lorenz-Mie scattering coefficients $b_n$ are determined by means of the proper boundary conditions established at $r = R$, which are continuity of tangential components of the electromagnetic field.
For illustrative purposes, let us consider a perfect electric conductor at the boundary.
Therefore, the tangential component of $\mathbf{E}_{tot}$, which is oriented along the $y$ axis, must vanishes at $r = R$.
This finally yields
\begin{equation}
b_n = - 2 \pi a_n \frac{J_n (k R)}{H_n^{(1)} (k R)},
\end{equation}
where the coefficients with different index $n$ can be treated separately due to the linear independence of the multipolar components of the wave field.

\section{Practical application of the series expansion}

The Bessel wave expansion can be applied to well-known patterns to confirm the validity of our approach.
First we consider a plane wave propagating along the $z$-axis, where the normalized field $E_{PW}(x,0) = 1$ is uniform on the ``focal'' plane.
In this case, the far field wave function $a_F (\theta) = \delta (\theta)$, corresponding to the Dirac delta function.
Therefore, the Fourier coefficients take the constant value $a_n = 1/2 \pi$ apart from the index $n$, which in addition is consistent with the result attained by using Eq.~(\ref{eq11}).
Finally, the wave field expansion of $E_{PW}(x,z) = \exp (i k z)$ can be set as
\begin{equation}
 E_{PW}(r,\theta) = \sum_{n = -\infty}^{\infty} i^n J_n (k r) \exp(i n \theta) .
\end{equation}
Note that such expression can also be determined by using the Jacobi-Anger identity as commonly carried out in scattering problems in two dimensions \cite{Bohren98}.
It is noteworthy that precisely a plane wave requiring an infinite number of terms in the series expansion provides, on the other hand, an exact solution to the Helmholtz equation.

\subsection{Gaussian beams}

A Gaussian beam in two dimensions can be described as an analytical solution to the paraxial wave equation.
Provided that the normalized wave field at the plane of the beam waist, that is the focal plane, is given by $E_{GB}(x,0) = \exp \left( - x^2 / w_0^2 \right)$, where $w_0$ is the Gaussian beam width, the field distribution in the $xz$ plane is then evaluated as \cite{Siegman86}
\begin{equation}
 E_{GB}(x,z) = \sqrt{\frac{q(0)}{q(z)}} \exp \left[i k \left( z + \frac{x^2}{2 q(z)} \right) \right] ,
\end{equation}
where the complex radius $q(z) = z - i z_R$ and the Rayleigh range $z_R = \pi w_0^2 / \lambda$.
The Fraunhofer pattern can be observed in the far field region $|z| \to \infty$ where 
\begin{equation}
 E_{GB}(x,z) \to \sqrt{\frac{z_R}{i z}} \exp \left[ i k \left(z + \frac{x^2}{2 z} \right) \right]  \exp \left[ - \left( \frac{x/z}{w_0/z_R} \right)^2 \right] .
\end{equation}
Therefore, the angular spectrum of the Gaussian beam yields $a_F(\theta) = 1 / ( \sqrt{\pi} \theta_0 ) \exp \left( - {\theta^2}/{\theta_0^2} \right)$
which also shows a Gaussian distribution with angular width given by the so-called far-field beam angle $\theta_0 = \lambda / \pi w_0$.
Finally, the Fourier coefficients of the series expansion are then written as $a_n = 1 / 2 \pi \exp \left( -n^2 \theta_0^2 / 4 \right)$.
Note that the resulting wave expansion of the Gaussian beam by applying our procedure is analogous to that found in Ref.~\cite{Kozaki82} following a completely different approach.

\begin{figure}[tb]
\centering
\includegraphics[width=\linewidth]{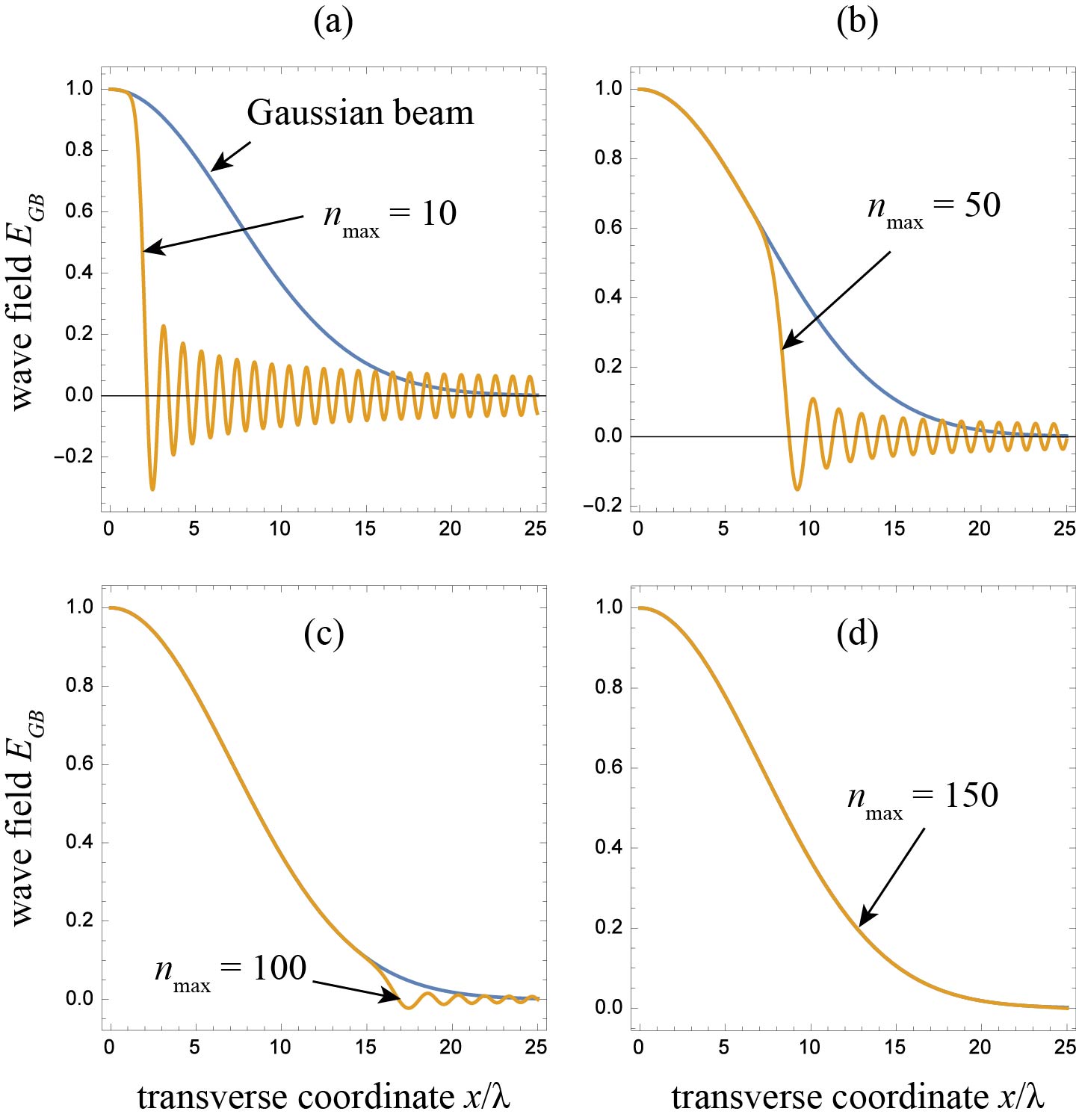}
\caption{Wave field $E_{GB}$ of a Gaussian beam evaluated at the beam waist plane $z=0$ compared with the series expansion given in Eq.~(\ref{eq13}), provided that the sequence is truncated with a maximum value of the index $n$ chosen as: (a) $n_\mathrm{max} = 10$, (b) $n_\mathrm{max} = 50$, (c) $n_\mathrm{max} = 100$, and (d) $n_\mathrm{max} = 150$.
	 The beam width is $w_0 = 10 \lambda$.}
\label{fig01}
\end{figure}

Figure~\ref{fig01} shows the field distribution of a Gaussian beam of waist $w_0 = 10 \lambda$, which corresponds to a far-field beam angle $\theta_0 = 1.8^\circ$, evaluated in the plane of the beam waist $z=0$.
For the sake of comparison, we use the series expansion given in Eq.~(\ref{eq13}) which is truncated by a finite Bessel order, $n \le n_\mathrm{max}$, the latter here called \emph{truncation index}. 
When $n_\mathrm{max} = 10$, the wave field is accurately evaluated only in the vicinities of the focal point, and deviations of the truncated series expansion are evident if $|x| > 1.5 \lambda$, as shown in Fig.~\ref{fig01}(a).
This result can be explained by taking into account that the Bessel wave field $J_n(k r)$ features a central circular shadow bounded by a main peak set at a distance roughly estimated as $r_n = |n|/k$, together with ripples dispersed further from focus \cite{Naserpour15b}.
As a consequence, the Bessel expansion of Eq.~(\ref{eq13}) which is subsequently truncated at $n_\mathrm{max}$ provides a rigorous estimation of the wave field at points of the focal region set well inside the circle of radius ${n_\mathrm{max}}/k$.
For $n_\mathrm{max} = 10$, $50$, $100$, and $150$ we estimate a boundary of radius $1.6 \lambda$, $8.0 \lambda$, $15.9 \lambda$, and $23.9 \lambda$ respectively, which is in good agreement with the results displayed in Figs.~\ref{fig01}(a)-(d).
Since the focal field of the Gaussian beam is negligible at $|x| > 3 w_0$, we finally conclude that $n_\mathrm{max} = 3 k w_0$ (that is $n_\mathrm{max} = 6/ \theta_0$) yields a sufficient number of Bessel wave functions to be used in the series expansion.
Note that for $w_0 = 10 \lambda$ we estimate that a number of elements higher than $n_\mathrm{max} = 60 \pi \approx 188$ in the Bessel expansion will not improve significantly the evaluation of the wave field in the focal plane.

\begin{figure}[tb]
	\centering
	\includegraphics[width=.8\linewidth]{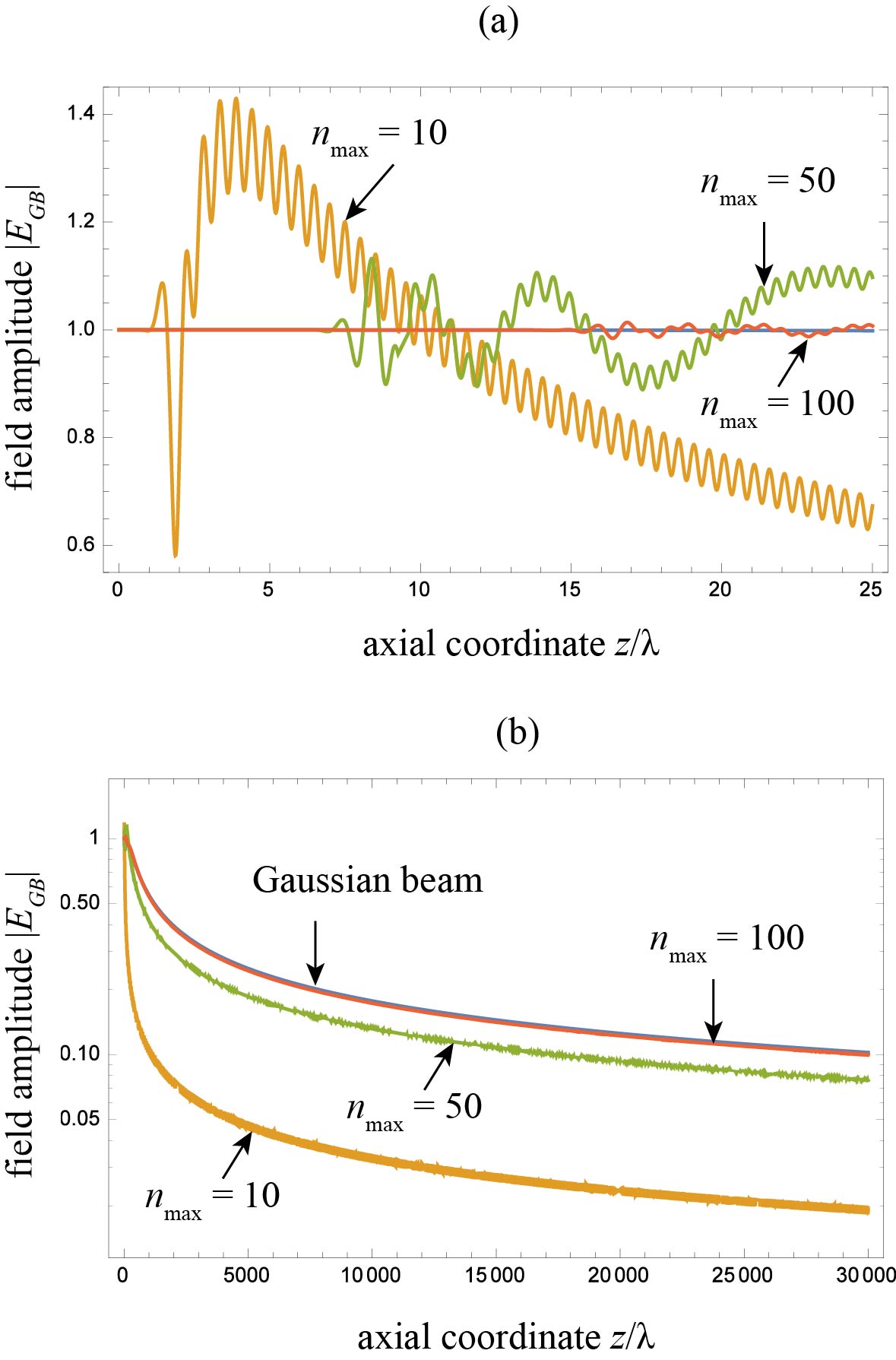}
	\caption{(a) Magnitude of the field $|E|$ evaluated by means of the series expansion (\ref{eq13}) along the optical axis of a Gaussian beam of width $w_0 = 10 \lambda$.
		We also include the exact on-axis field distribution derived from the paraxial wave equation.
		(b) Log plot of the same calculation which is performed in a domain spread out well beyond the Rayleigh range $z_R = 314 \lambda$.}
	\label{fig02}
\end{figure}

Inasmuch as the Bessel wave functions constituting the series expansion exhibit radial symmetry, it is reasonable to identify equivalent restrictions on the sequence truncation when evaluating the field distribution of the Gaussian beam along the optical axis.
The latter is proved in Fig.~\ref{fig02}(a), which shows the magnitude of the series expansion at $x=0$ for a Gaussian beam of $w_0 = 10 \lambda$, considering a different number of Bessel wave functions used in the truncated sequence.
The exact value $|E_{GB}(0,z)| = \left[ 1 + (z/z_R)^2 \right]^{-1/4}$ is also represented for the sake of comparison \cite{Siegman86}.
Due to the slower decay of the Gaussian field distribution along the optical axis, which is characterized by the Rayleigh range ($z_R = 314 \lambda$ is much longer than the beam width), it might occur that an accurate computation of the on-axis field distribution using the Bessel series expansion, at points reaching significant values, would require a higher number $n_\mathrm{max}$.
To verify the validity of this argument, Fig.~\ref{fig02}(b) shows a semi-log plot of the magnitude derived from the series expansion from focus to a point located well beyond the Rayleigh range.
Though using $n_\mathrm{max} = 100$ it is expected a significant deviation of our calculation moving away to a distance further than roughly $16 \lambda$, we surprisingly obtain a tolerable estimation up to $3 \times 10^4 \lambda$.
The strong sidelobes featuring the Bessel function also play a key role in the determination of the wave field, particularly on the optical axis.
We conclude that the truncation index $n_\mathrm{max}$ indicating the number of elements used in the truncated series expansion does not strictly determines the region around the focus where the Eq.~(\ref{eq13}) is applicable but the working precision of our calculation.

\begin{figure}[tb]
	\centering
	\includegraphics[width=.8\linewidth]{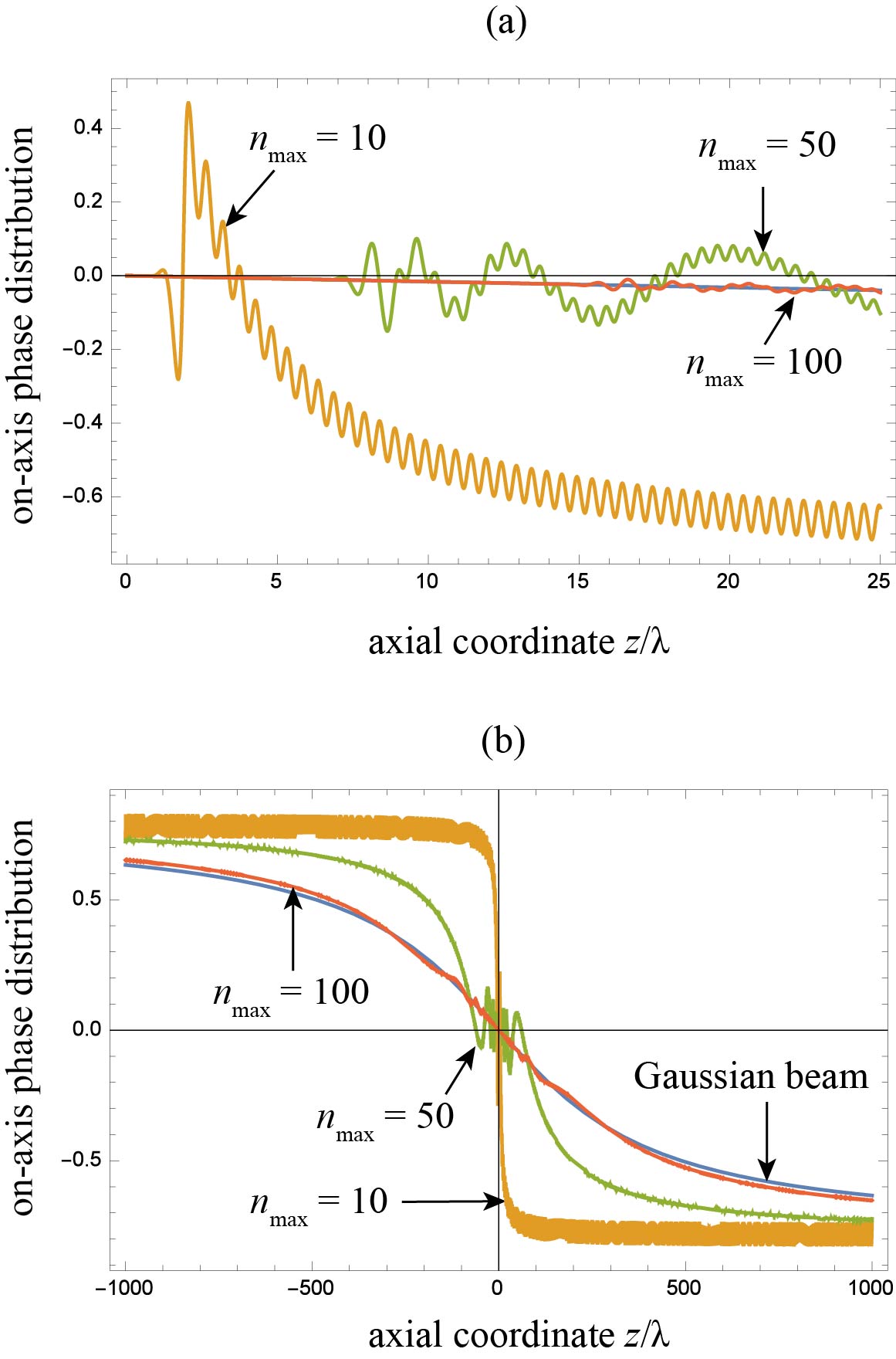}
	\caption{The same as in Fig.~\ref{fig02} but showing the argument of the wave field.
		The term $\exp(i k z)$ of the wave field is disregarded, which leads to fast-evolving variations of the phase distribution.}
	\label{fig04}
\end{figure}

The above discussion can include the analysis of the phase distribution of the Gaussian beam along its optical axis, as shown in Fig.~\ref{fig04}(a) and (b).
Note that the exact phase distribution is given by $-\phi(z)/2$, where $\tan \phi = z/z_R$, provided that the fast-evolving term $\exp(i k z)$ is ignored for the sake of clarity \cite{Siegman86}.
The Gouy phase shift accumulated when passing through the focus from $z \to -\infty$ to $z \to +\infty$ is $-\pi/2$, a fact that is primarily carried out in the range $|z| < z_R$.
As evidenced in Fig.~\ref{fig04}(b), the series expansion (\ref{eq13}) always provides the correct Gouy phase shift independently of the applied truncation. 
However, underestimating the value of $n_\mathrm{max}$ will reduce the path wherein such a phase shift is naturally induced.

\begin{figure*}[h]
	\centering
	\includegraphics[width=.95\linewidth]{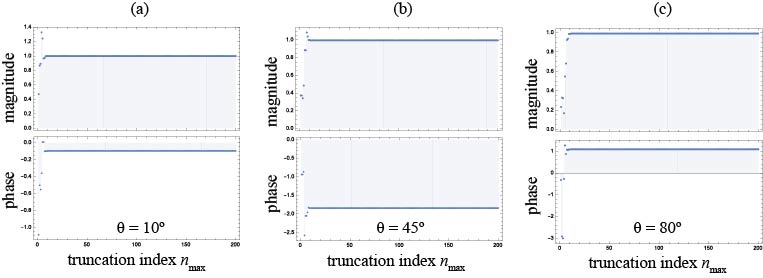}
	\caption{Convergence analysis of the series expansion (\ref{eq13}) at varying truncation index $n_\mathrm{max}$, analyzed at three different points equidistant from focus ($r = \lambda$) in the first quadrant of the $xz$ plane: $(x,z) = r (\sin \theta, \cos \theta) = $ (a) $(0.17,0.98) \lambda$, (b) $(0.71,0.71) \lambda$, and (c) $(0.98,0.17) \lambda$. 
		The Gaussian beam width is again $w_0 = 10 \lambda$.}
	\label{fig05_1}
\end{figure*}

\begin{figure*}[h]
	\centering
	\includegraphics[width=.95\linewidth]{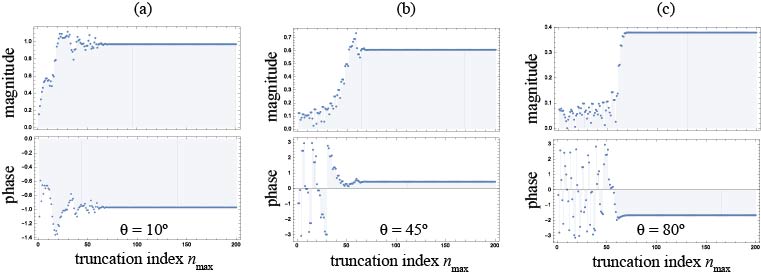}
	\caption{The same as in Fig.~\ref{fig05_1} but the three points are further from focus at a distance $r = 10 \lambda$: $(x,z) = r (\sin \theta, \cos \theta) = $ (a) $(1.7,9.8) \lambda$, (b) $(7.1,7.1) \lambda$, and (c) $(9.8,1.7) \lambda$.}
	\label{fig05_2}
\end{figure*}

\begin{figure*}[h]
	\centering
	\includegraphics[width=.95\linewidth]{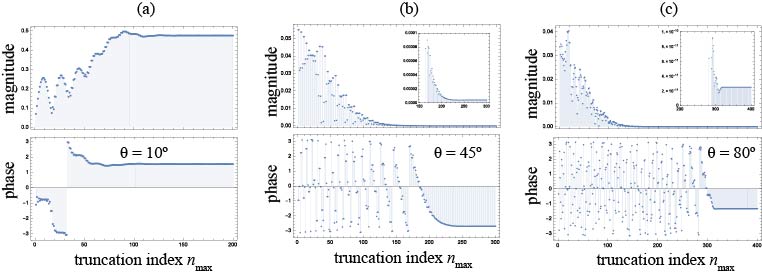}
	\caption{The same as in Fig.~\ref{fig05_1} but the three points are further from focus at a distance $r = 50 \lambda$: $(x,z) = r (\sin \theta, \cos \theta) = $ (a) $(8.7,49) \lambda$, (b) $(35,35) \lambda$, and (c) $(49,8.7) \lambda$.}
	\label{fig05_3}
\end{figure*}

Figures~\ref{fig05_1}-\ref{fig05_3} illustrate the convergence of the series expansion (\ref{eq13}), by varying the truncation index $n_\mathrm{max}$, which is evaluated at different points of the first quadrant in the $xz$ plane.
The chosen points have one of these three different radii: $r = \{\lambda, 10 \lambda, 50 \lambda \}$, given at three different angular orientations, $\theta = 10^\circ$, $45^\circ$, and $80^\circ$.
For each one of these nine points we represent the magnitude and phase of the complex wave field separately.
When the computation of the field is performed close to focus, at $r=\lambda$, a fast convergence of the series expansion is reached where $n_\mathrm{max} = 10$, which is a few integers higher than the product $k r$, is enough to accurately determine the complex field.
Such behavior does not depend on the azimuthal angle $\theta$, as shown in Figs.~\ref{fig05_1}(a)-(c).
By analyzing the points set at $r=10\lambda$, where now the magnitude of the field drops more significantly (yet remains in the same order) at higher values of $\theta$, that is closer to the focal plane than the optical axis, the series expansion on the other hand converges isotropically.
In Fig.~\ref{fig05_2} we recognize that a truncation index $n_\mathrm{max} = 70$ is appropriate for all angles; note that here $k r \approx 63$.

However, a dramatic variation of the series convergence occurs if we draw attention to points at a distance $r=50\lambda$ far from focus, as shown in Fig.~\ref{fig05_3}.
In this case, $n_\mathrm{max} = 100$ suffices for the point set near to the optical axis at $\theta = 10^\circ$; on the contrary the truncation index $n_\mathrm{max}$ should surpass a value of $300$ when $\theta = 80^\circ$ in order to achieve an accurate estimation of the wave field.
Importantly, the magnitude of the Gaussian field $|E_{GB}|$ reaches a value of $0.48$ in the first case, falling to $|E_{GB}| = 2.5 \times 10^{-11}$ in the latter, which is a consequence of the natural resistance of paraxial fields to decay along the direction of propagation.
Therefore, the requirement establishing a truncation index lightly higher than the parameter $k r$ ($\approx 314$ here) is greatly moderated near the optical axis where the field reaches its highest values by several orders of magnitude.

\subsection{Apertured focused fields}

Let us consider an apertured focal wave that can be produced for instance by a lens system, where in addition pupil truncation is taken into account.
In this case, the angular spectrum might be assumed to be flat, which is expressed as $a_F(\theta) = a_F(0)$ within the interval $|\theta| < \theta_0$; otherwise the angular spectrum of the wave field vanishes.
Note that $\theta_0$ represents the numerical aperture of the converging wave in the paraxial regime, provided that propagation takes place in free space.
The wave field in the focal volume can be evaluated by means of the Debye diffraction integral given in Eq.~(\ref{eq14}), which yields
\begin{eqnarray}
 E_{AF} &=& \frac{1}{2 \sqrt{\zeta}} \exp \left[i k \left(z + \frac{x^2}{2 z}\right)\right] \label{eq15} \\
      && \times \left\{\tilde{F}^*\left[\sqrt{\zeta} \left( 1-\frac{\theta}{\theta_0}\right)\right] + \tilde{F}^*\left[\sqrt{\zeta} \left( 1 + \frac{\theta}{\theta_0} \right) \right] \right\} , \nonumber
\end{eqnarray}
where the normalized spatial coordinate $\zeta = 2 \theta_0^2 z / \lambda$ and $\tilde{F}^*(\cdot)$ is the complex conjugate of the complex Fresnel integral function \cite{Siegman86}.
Also we used $a_F(0) = 1/2 \theta_0$ to obtain a normalized in-focus wave field $E_0 = 1$.
Specifically at the focal plane $z=0$, the wave field reduces to
\begin{equation}
 E_{AF}(x,0) = \frac{\sin \left( k \theta_0 x \right)}{k \theta_0 x} .
\end{equation}
The first zero of the wave field which is found at $x_1$ around the focal point satisfies $k \theta_0 x_1 = \pi$ and determines the limit of resolution following the Rayleigh criterion.
If we conserve the definition of the beam width previously used for Gaussian beams ($w_0 = \lambda/\pi \theta_0$) as measured at the focal plane, we may write $x_1 = (\pi / 2) w_0 \approx 1.57 w_0$.
Therefore, here $w_0$ can also be considered as the beam width of the apertured focused wave.
Finally, the Fourier coefficients to be employed in the series expansion (\ref{eq13}) now read as $a_n = {\sin \left( n \theta_0 \right)}/{\left(2 \pi n \theta_0 \right)}$.

\begin{figure}[tb]
	\centering
	\includegraphics[width=\linewidth]{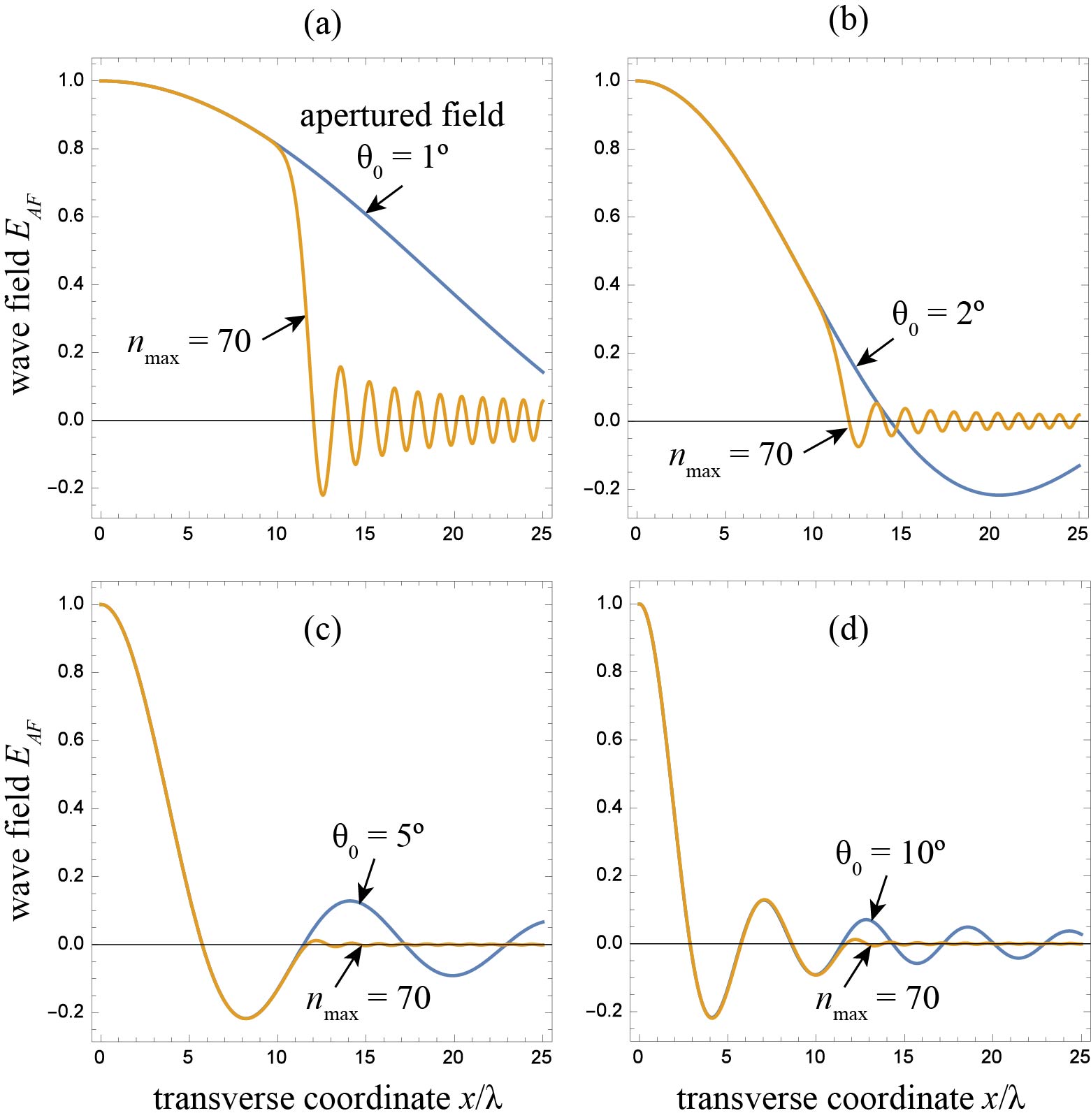}
	\caption{Wave field $E_{AF}$ of an apertured wave field evaluated at the focal plane $z=0$ compared with the series expansion (\ref{eq13}) which is characterized by a truncation index $n_\mathrm{max} = 70$, for different numerical apertures: (a) $\theta_0 = 1^\circ$, (b) $\theta_0 = 2^\circ$, (c) $\theta_0 = 5^\circ$, and (d) $\theta_0 = 10^\circ$.}
	\label{fig06}
\end{figure}

The focal field of the apertured converging wave is represented in Fig.~\ref{fig06} for different numerical apertures $\theta_0$, together with its corresponding Bessel series expansion which is truncated at $n_\mathrm{max} = 70$.
The truncated series expansion proves an accurate evaluation of the wave field from focus to a distance $~11 \lambda$, which on the other hand is roughly the estimator $n_\mathrm{max} / k$.
When the numerical aperture is low, it may happen that the region of validity of the series expansion even cannot reach the central lobe of the \emph{sinc} pattern, as illustrated in Fig.~\ref{fig06}(a) and (b) for $\theta_0 = 1^\circ$ and $\theta_0 = 2^\circ$, respectively. 
This peak of highest intensity will be conveniently reproduced by the series expansion (\ref{eq13}) provided that $n_\mathrm{max} > k x_1$, occurring if $\theta_0 > \pi / n_\mathrm{max}$ ($\theta_0 > 2.6^\circ$ for $n_\mathrm{max} = 70$).
In Fig.~\ref{fig06}(c) we consider the increased numerical aperture $\theta_0 = 5^\circ$, which associated Bessel expansion covers not only the central lobe but the first sidelobe accurately.
These sidelobes are determined by the zeros of the sine function found at $x_m = m / k \theta_0$, thus a given truncation index ensures an accurate reproduction of the wave field containing the central peak and $m-1$ sidelobes determined by $m=n_\mathrm{max} \theta_0 / \pi$.
Specifically, for $\theta = 5^\circ$ we appropriately represent up to $m-1 \approx 1$ sidelobes, whereas such quantity is tripled for $\theta = 10^\circ$, as shown in Figs.~\ref{fig06}(c) and (d).

\begin{figure}[tb]
	\centering
	\includegraphics[width=\linewidth]{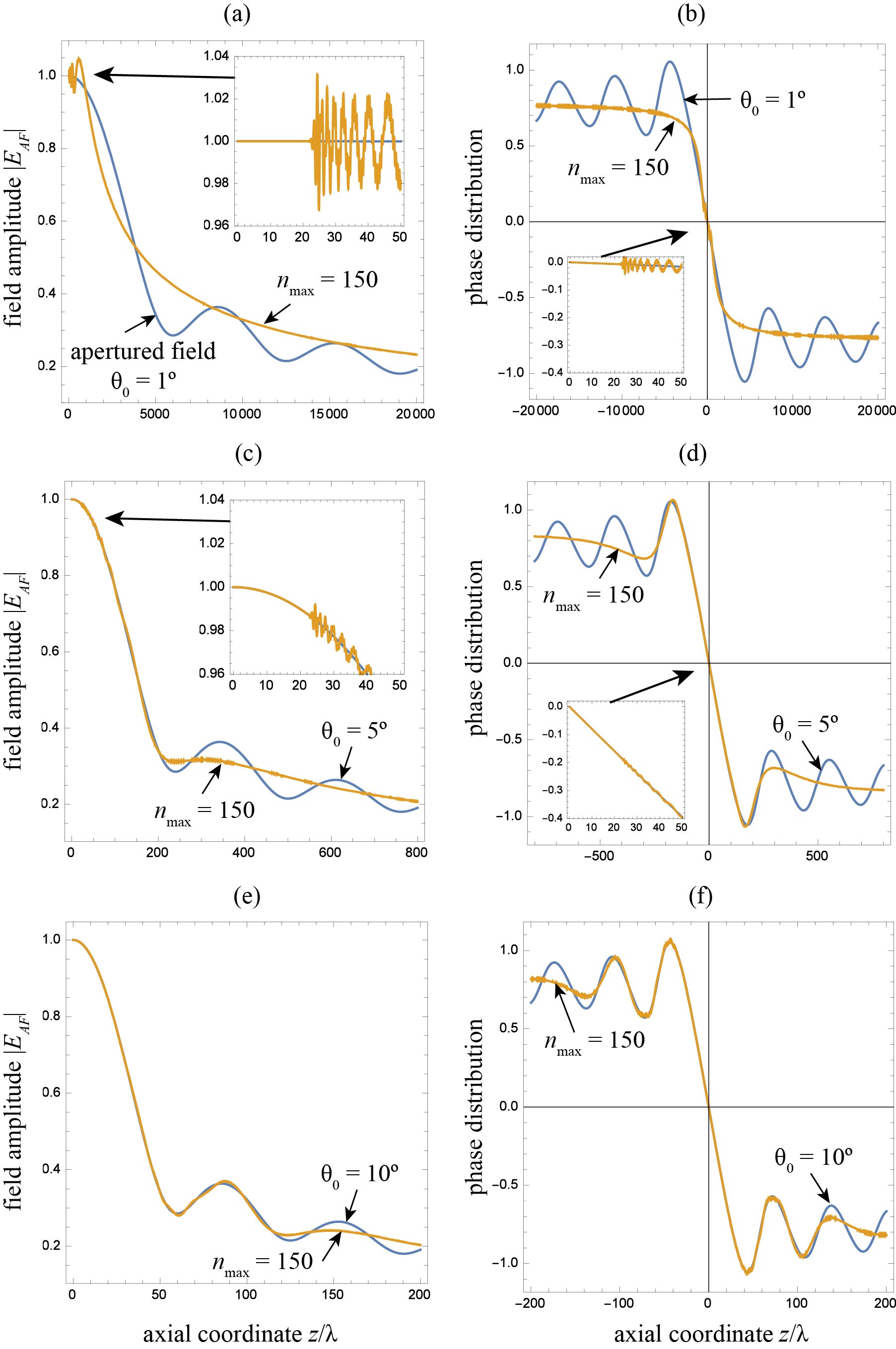}
	\caption{Magnitude and phase of the on-axis complex field distribution of an apertured focused wave, evaluated by means of the exact paraxial equation (\ref{eq15}) and by the series expansion (\ref{eq13}) for the numerical apertures: $\theta_0 = 1^\circ$ in (a) and (b), $\theta_0 = 5^\circ$ in (c) and (d), and $\theta_0 = 10^\circ$ in (e) and (f).}
	\label{fig07}
\end{figure}

The analysis of the on-axis field distribution is illustrated in Fig.~\ref{fig07}.
The complex wave field depends on the normalized spatial coordinate $\zeta$ exclusively, leading to a magnification of the pattern in direct proportion to $1/\theta_0^2$, in opposition to the transverse pattern whose magnification depends on $1/\theta_0$.
When the series expansion is truncated at $n_\mathrm{max} = 150$, accurate estimations are derived within the interval ranged from focus to roughly $20 \lambda$ (note that $n_\mathrm{max}/k = 23.87 \lambda$).
Out of that range fast oscillations in phase and magnitude make the series expansion deviates from the exact on-axis field distribution, as evidenced in Fig.~\ref{fig07}(a)-(d).
However, the amplitude of such oscillations are notably reduced when the numerical aperture $\theta_0$ increases.
Nevertheless, an adequate field decay at much longer distances as $1/\sqrt{z}$ is reproduced by the series expansion even at the lowest numerical aperture shown in Fig.~\ref{fig07}(a).
The exact phase distribution now exhibits a strong linear variation, conjoined with the central peak of intensity, along with some maxima and minima around the asymptotes $\pm \pi/4$.
The series expansion provides a good estimation of the slope in the central linear region even in the case of $\theta_0 =1^\circ$ where $n_\mathrm{max} = 150$ remains suboptimal to accurately mimic the whole central peak of the field pattern.
The asymptotes are also well established in the latter case.
By increasing the value of the angle $\theta_0$ we can in addition follow appropriately the profile of the continuous side lobes that appear when we move further from focus. 
When $\theta_0 = 10^\circ$, the central peak and the first sidelobe of the intensity profile are conveniently determined by the series expansion, as well as the focal linear variation of the phase distribution and its three closest lateral extrema, as shown in Fig.~\ref{fig07}(e) and (f).

\section{Conclusions}

Based on the far field characteristics of paraxial beams, we are able to represent the full wave field in a series expansion in terms of Bessel wave functions which are solutions of the Helmholtz equation in two dimensions.
The coefficients of the series expansion in fact stem from a Fourier expansion of the wave-field angular spectrum, and they can be analytically determined by using the pattern in the focal plane wherein the field is mostly localized.
This representation of the wave field is adequate for scattering problems where a scatterer is localized in a circular section around the origin of coordinates, as described thoroughly.

Application of the Bessel expansion to a uniform plane wave leads us to one expression which, alternatively, may be inferred from the well-known Jacobi-Anger identity, demonstrating the validity of our approach.
Other paraxial beams have been analyzed in detail.
Firstly a Gaussian beam has been critically examined, particularly in comparison with a practical truncation given by the index $n_\mathrm{max}$ of the corresponding series expansion.
The wave field is accurately evaluated at points whose distance from the origin of coordinates is shorter that $n_\mathrm{max}/k$.
Provided that such origin of coordinates is established at the maximum of intensity, then a truncation index $n_\mathrm{max} = 3 k w_0$ (where $w_0$ is the beam waist) should be enough for the series expansion to provide a good estimation of the wave field in the whole space.

By analyzing apertured wave fields we further introduce the main aspects of the series truncation.
For instance, intensity sidelobes existing in the vicinity of the focal peak are critically reproduced under the appropriate truncation index.
Due to the specific characteristics of the paraxial beams, such a circumstance is much more evident transversally to its direction of propagation.
Nevertheless, both the inverse root squared decay of the field with distance from focus, and the $\pi/2$ phase shift undergone on axis for general wave fields is always implicit in the Bessel-based series expansion.
Finally, this procedure can be applied to any kind of two-dimensional paraxial beam, in particular when it will interact with particles set in its trajectory.
Such scatterers are not restricted in size, ranging from nanowires to large cylinders, which in addition can be made of anisotropic and even multilayered materials \cite{Diaz16d,Diaz17}.

\section*{Acknowledgments}

This work was supported by the Spanish Ministry of Economy and Competitiveness (MINECO) (TEC2014-53727-C2-1-R).

\bibliographystyle{elsarticle-num}

\end{document}